\documentclass[apjl]{emulateapj}
\usepackage{natbib}

\newcommand{\teff}{\mbox{$T_{\rm eff}$}}
\newcommand{\logg}{\mbox{$\log g$}}
\newcommand{\vsini}{\mbox{$v \sin i$}}
\newcommand{\mictrb}{\mbox{$\xi_{\rm t}$}}
\newcommand{\mactrb}{\mbox{$v_{\rm mac}$}}

\newcommand{\kms}{\mbox{km\, s$^{-1}$}}
\newcommand{\halpha}{\mbox{$H_\alpha$}}
\newcommand{\hbeta}{\mbox{$H_\beta$}}

\shorttitle{The transiting exoplanet WASP-15b}
\shortauthors{West et al.}

\begin{document}

\title{The low density transiting exoplanet WASP-15b}

\author{R.~G.~West\altaffilmark{1}}
\author{D.~R.~Anderson\altaffilmark{2}}
\author{M.~Gillon\altaffilmark{3,4}}
\author{L.~Hebb\altaffilmark{5}}
\author{C.~Hellier\altaffilmark{2}}
\author{P.~F.~L.~Maxted\altaffilmark{2}}
\author{D.~Queloz\altaffilmark{3}}
\author{B.~Smalley\altaffilmark{2}}
\author{A.~H.~M.~J.~Triaud\altaffilmark{3}}
\author{D.~M.~Wilson\altaffilmark{2}}
\author{S.~J.~Bentley\altaffilmark{2}}
\author{A.~Collier~Cameron\altaffilmark{5}}
\author{B.~Enoch\altaffilmark{5}}
\author{K.~Horne\altaffilmark{5}}
\author{J.~Irwin\altaffilmark{6}}
\author{T.~A.~Lister\altaffilmark{7}}
\author{M.~Mayor\altaffilmark{3}}
\author{N.~Parley\altaffilmark{5}}
\author{F.~Pepe\altaffilmark{3}}
\author{D.~Pollacco\altaffilmark{8}}
\author{D.~Segransan\altaffilmark{3}}
\author{S.~Udry\altaffilmark{3}}
\author{P.~J.~Wheatley\altaffilmark{9}}

\altaffiltext{1}{Department of Physics and Astronomy, University of Leicester, Leicester, LE1 7RH, UK}
\altaffiltext{2}{Astrophysics Group, Keele University, Staffordshire, ST5 5BG, UK}
\altaffiltext{3}{Observatoire de Gen\`eve, Universit\'e de Gen\`eve, 51 Chemin des Maillettes, 1290 Sauverny, Switzerland}
\altaffiltext{4}{Institut d'Astrophysique et de G\'eophysique,  Universit\'e de Li\`ege,  All\'ee du 6 Ao\^ut, 17,  Bat.  B5C, Li\`ege 1, Belgium}
\altaffiltext{5}{School of Physics and Astronomy, University of St. Andrews, North Haugh, Fife, KY16 9SS, UK}
\altaffiltext{6}{Department of Astronomy, Harvard University, 60 Garden Street, MS 10, Cambridge, Massachusetts 02138, USA}
\altaffiltext{7}{Las Cumbres Observatory, 6740 Cortona Dr. Suite 102, Santa Barbara, CA 93117, USA}
\altaffiltext{8}{Astrophysics Research Centre, School of Mathematics \& Physics, Queen's University, University Road, Belfast, BT7 1NN, UK}
\altaffiltext{9}{Department of Physics, University of Warwick, Coventry, CV4 7AL, UK}

\begin{abstract}
  We report the discovery of a low-density exoplanet transiting an 11th
  magnitude star in the Southern hemisphere. WASP-15b, which orbits its host
  star with a period $P=3.7520656\pm0.0000028\,$d has a mass $M_{\rm p}=0.542\pm
  0.050\,$M$_{\rm J}$ and radius $R_{\rm p}=1.428\pm 0.077\,$R$_{\rm J}$, and is
  therefore the one of least dense transiting exoplanets so far discovered
  ($\rho_p=0.247\pm 0.035\,$g\,cm$^{-3}$). An analysis of the spectrum of the
  host star shows it to be of spectral type around F5, with an effective
  temperature $T_{\rm eff}=6300\pm100\,$K and $[{\rm
    Fe/H}]=-0.17\pm 0.11$.
\end{abstract}

\section{Introduction}

Transiting exoplanets represent the best current opportunity to test
theoretical models of the internal structure of such planets, and the
formation and evolution of planetary systems. At the time of writing the
discovery of approaching sixty transiting systems had been announced in the
literature by numerous well-established survey projects, such as HATnet
\citep{bakos04}, XO \citep{mccullough05}, TrES \citep{odonovan06} and WASP
\citep{pollacco06}.

The WASP project operates two identical observatories, one at La Palma in the
Canary Islands, and the other at Sutherland in South Africa. Each telescope
has a field of view of approximately 500 square degrees. The WASP survey is
capable of detecting planetary transit signatures in the light-curves of hosts
in the magnitude range V\,$\sim$9--13. A full description of the telescope
hardware, observing strategy and pipeline data analysis is given in
\citet{pollacco06}.

\section{Observations}
The host star WASP-15 (= 1SWASP~J135542.70-320934.6 = 2MASS~13554269-3209347 =
USNO-B1.0~0578-0402627 = NOMAD1~0578-0409366 = TYCH2~7283-01162-1) is
catalogued as a star of magnitude $V=11.0$ and co-ordinates $\alpha=13^{\rm
  h}55^{\rm m}42\fs 71$, $\delta=-32\degr 09\arcmin 34\farcs 6$. WASP-15 was
observed by the WASP-South observatory in a single camera field from
2006~May~04 to 2006~July~17, and in two overlapping camera fields from
2007~January~31 to 2007~July~17 and from 2008~January~31 to 2008~May~29.

\begin{figure}
\plotone{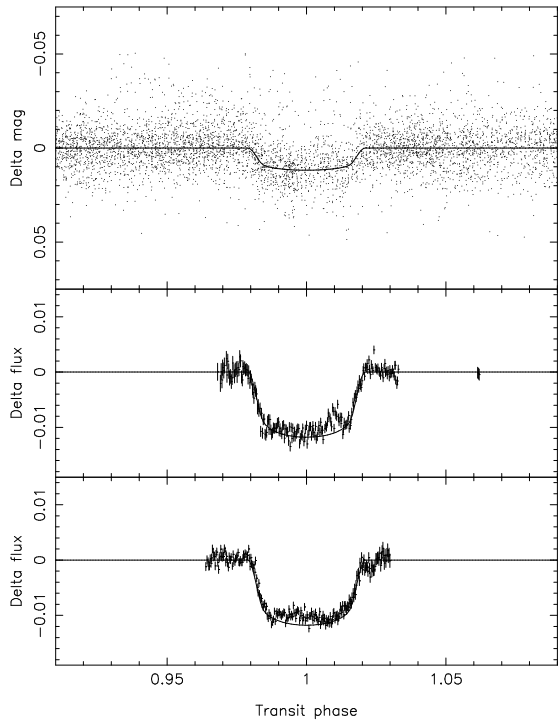}
\caption{WASP photometry folded on the best-fit period (top panel). EULER I-band
  (middle) and R-band (lower). Curve shows the best-fit transit model from the
  MCMC fitting.}
\label{foldlc}
\end{figure}

The data were processed using the project's routine analysis pipeline,
de-trending and transit-detection tools as described in \cite{pollacco06},
\cite{cameron06} and \cite{cameron07}. A total of 24943 data points were
acquired, in which was detected a recurrent transit signature with a period of
3.7520 days and a depth of $~0.011$ mag (Fig.~\ref{foldlc}, top panel). In
total some 11 full or partial transits were observed by WASP-South.

Follow-up photometric observations were made using the EulerCAM photometer on
the 1.2m Euler telescope in the I-band on 2008~March~29 and the R-band on
2008~May~13 (Fig.~\ref{foldlc} middle and lower panel), which confirmed the
presence of a flat-bottomed dip expected from the transit of an exoplanet.
Both transit light-curves from EulerCAM exhibit excess variability likely due
to systematic noise.

\begin{figure}
\plotone{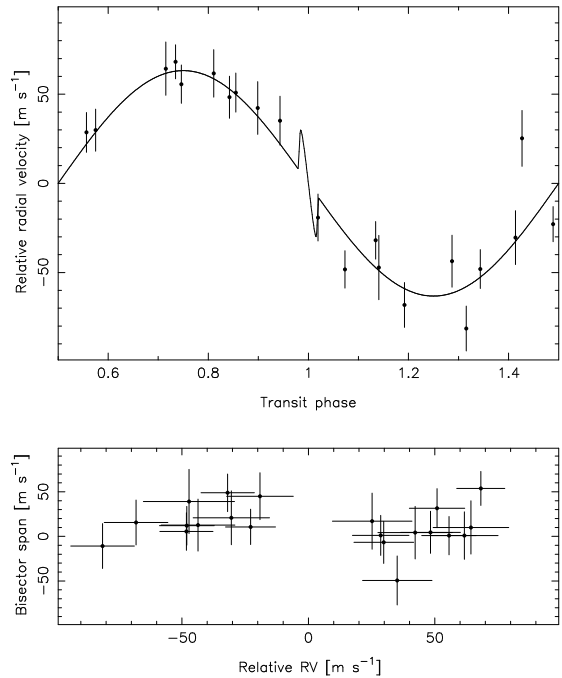}
\caption{CORALIE radial-velocity measurements. Curve shows best-fit MCMC
  model. Lower panel shows the bisector span plotted against RV.}
\label{rvplot}
\end{figure}

Subsequent observations using the CORALIE spectrograph on the Euler telescope
between 2008~March~06 and 2008~July~17 yielded 21 radial-velocity measurements
(Table~\ref{rvdata}; Fig.~\ref{rvplot} upper panel) which show a sinusoidal
variation with a semi-amplitude of around 65\,m\,s$^{-1}$ on the same period
as the transit signature. An analysis of the bisector spans (Fig.~\ref{rvplot}
lower panel) shows no correlation with the measured radial velocity, which
rules out the possibility that the RV variations were due to a blended
eclipsing binary system.

\begin{deluxetable}{lrrr}
\tablewidth{0pt}
\tablecaption{Radial velocity measurements of WASP-15}
\tablehead{
\colhead{BJD} & \colhead{RV} & \colhead{$\sigma_{RV}$} & \colhead{BS}\\
\colhead{$-2\,400\,000$} & \colhead{(km s$^{-1}$)} & \colhead{(km s$^{-1}$)} &
\colhead{(km s$^{-1}$)}
}
\startdata
54531.8146 & $-2.3020$ & 0.0148 & 0.0042\\
54532.7221 & $-2.3914$ & 0.0180 & 0.0392\\
54533.7468 & $-2.3747$ & 0.0151 & 0.0209\\
54534.8778 & $-2.2799$ & 0.0150 & 0.0100\\
54535.7341 & $-2.3090$ & 0.0137 & $-0.0494$\\
54536.6666 & $-2.4124$ & 0.0126 & 0.0156\\
54537.7805 & $-2.3671$ & 0.0099 & 0.0107\\
54538.7459 & $-2.2886$ & 0.0108 & 0.0009\\
54556.7948 & $-2.3156$ & 0.0112 & 0.0011\\
54557.7488 & $-2.2825$ & 0.0133 & 0.0010\\
54558.7334 & $-2.3925$ & 0.0105 & 0.0055\\
54559.7479 & $-2.3923$ & 0.0109 & 0.0120\\
54560.6158 & $-2.3144$ & 0.0118 & $-0.0067$\\
54589.6590 & $-2.4257$ & 0.0126 & $-0.0108$\\
54591.6355 & $-2.2959$ & 0.0118 & 0.0046\\
54655.4689 & $-2.2933$ & 0.0110 & 0.0316\\
54656.5165 & $-2.3761$ & 0.0106 & 0.0489\\
54657.6134 & $-2.3190$ & 0.0157 & 0.0171\\
54662.5205 & $-2.2760$ & 0.0095 & 0.0538\\
54663.5879 & $-2.3634$ & 0.0132 & 0.0451\\
54664.5932 & $-2.3879$ & 0.0146 & 0.0126\\
\enddata
\label{rvdata}
\end{deluxetable}

\section{Evolutionary status of the host star}

The individual CORALIE spectra are relatively low signal-to-noise, but when
co-added into 0.01\AA\ steps they give a S/N of around 80:1 which is suitable
for a photospheric analysis of WASP-15. In addition, a single HARPS spectrum was
used to complement the CORALIE analysis, but this spectrum had relatively modest
S/N of around 50:1.

An analysis of the available spectral data was performed using the {\sc
  uclsyn} spectral synthesis package \citep{smith92, smalley01} and {\sc
  atlas9} models without convective overshooting \citep{castelli97}. The
\halpha\ and \hbeta\ lines were used to determine the effective temperature
(\teff), while the Na {\sc i} D and Mg {\sc i} b lines were used as surface
gravity (\logg) diagnostics.  Additionally, the Ca H \& K lines provide a
further check on the derived \teff\ and \logg. This fit yielded a $T_{\rm
  eff}=6300\pm 100\,$K and $\log g=4.35\pm 0.15$ (Table~\ref{hostparams}).

In order to determine the elemental abundances the equivalent widths of
several clean and unblended lines were measured. Atomic line data were mainly
taken from the \citet{kurucz95} compilation, but with updated van der Waals
broadening coefficients for lines in \citet{barklem00} and $\log gf$ values
from \citet{gonzalez00}, \citet{gonzalez01} or \citet{santos04}. A value for
microturbulence (\mictrb) was determined from Fe~{\sc i} using the method of
\citet{magain84}. The ionization balance between Fe~{\sc i} and Fe~{\sc ii}
and the null-dependence of abundance on excitation potential were used in
addition to the \teff\ and \logg\ diagnostics \citep{smalley05}. The
abundances are given in Table~\ref{hostparams}. The quoted error estimates
include that given by the uncertainties in \teff, \logg\ and \mictrb, as well
as the scatter due to measurement and atomic data uncertainties. The Li {\sc
  i} 6708\AA\ line is not detected (EW $<$ 2m\AA), allowing us to derive an
upper-limit on the lithium abundance of log n(Li/H) + 12 $<$ 1.2. The \teff\ 
of this star implies it is in the lithium-gap \citep{bohm04}, so the lithium
abundance does not provide an age constraint.

Stellar rotation velocity (\vsini) was determined by fitting the profiles of
several unblended Fe~{\sc i} lines in the HARPS spectrum. A value for
macroturbulence (\mactrb) of 4.8~\kms\ was adopted, from the \citet{valenti05}
relationship, and an instrumental FWHM of $0.060\pm 0.005\,$\AA, determined
from the telluric lines around 6300\AA. A best fitting value of $v\sin i = 4
\pm 2\,$\kms\ was obtained.

\begin{figure}
\plotone{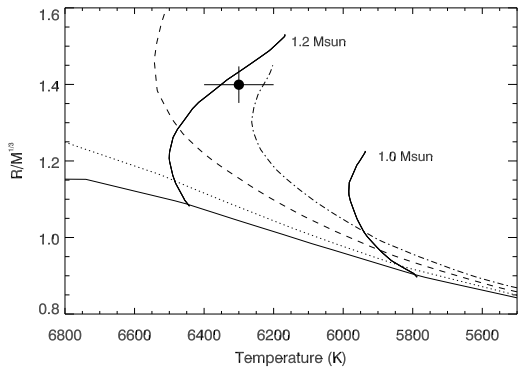}
\caption{The position of WASP-15 in the $R/M^{1/3}-$\teff\
  plane. Evolutionary tracks for a star of metallicity $[M/H]=-0.17$ from
  \citet{girardi00} are plotted along with isochrones for ages 100\,Myr
  (solid), 1\,Gyr (dotted), 2.5\,Gyr (dashed) and 4\,Gyr (dot-dashed).
  Evolutionary mass tracks are shown for 1.0 and 1.2\,M$_\odot$.}
\label{evolplot}
\end{figure}

To estimate the age of WASP-15 we compared the stellar density and temperature
measured from the photometric and spectroscopic analysis against the
evolutionary models of \citet{girardi00} interpolated to a metallicity
$[M/H]=-0.17$ (Fig.~\ref{evolplot}). This procedure yields a best-fit age of
$3.9^{+2.8}_{-1.3}\,$Gyr and a best-fit mass $M_*=1.19\pm0.10\,$M$_\sun$.

\begin{deluxetable}{lr}
\tablewidth{0pt}
\tablecaption{Parameters of the host star}
\tablehead{\colhead{Parameter} & \colhead{Value}}
\startdata
Stellar mass, $M_{\rm *}$ (M$_\sun$) & $1.18\pm 0.12$\\
Stellar radius, $R_{\rm *}$ (R$_\sun$) & $1.477\pm 0.072$\\
Stellar surface gravity, $\log g$\tablenotemark{a} (cgs) & $4.169\pm 0.033$\\
Stellar density, $\rho_{\rm *}$ ($\rho_\sun$) & $0.365\pm 0.037$\\
Stellar luminosity, $L_\star$ (L$_\sun$) & $3.09\pm 0.34$\\
Age (Gyr) & $3.9^{+2.8}_{-1.3}$ \\
\\
\teff (K)     & 6300 $\pm$ 100 \\
$\log g$\tablenotemark{b}      & 4.35 $\pm$ 0.15 \\
\mictrb\ (\kms)   & 1.4 $\pm$ 0.1  \\
\vsini\ (\kms)    & 4 $\pm$ 2  \\
{[Fe/H]}   &$-$0.17 $\pm$ 0.11 \\
{[Na/H]}   &$-$0.25 $\pm$ 0.05 \\
{[Mg/H]}   &$-$0.13 $\pm$ 0.12 \\
{[Si/H]}   &$-$0.15 $\pm$ 0.10 \\
{[Ca/H]}   &$-$0.06 $\pm$ 0.13 \\
{[Sc/H]}   &$-$0.07 $\pm$ 0.12 \\
{[Ti/H]}   &$-$0.14 $\pm$ 0.06 \\
{[V/H]}    &$-$0.20 $\pm$ 0.11 \\
{[Cr/H]}   &$-$0.11 $\pm$ 0.12 \\ 
{[Co/H]}   &$-$0.16 $\pm$ 0.08 \\
{[Ni/H]}   &$-$0.25 $\pm$ 0.08 \\
$\log {\rm N(Li)}$& $<$ 1.2 \\
%%\teff(IRFM) (K)&6400 $\pm$ 140\\
%%$\theta$(IRFM) (mas)& 0.046 $\pm$ 0.002 \\
\enddata
\tablenotetext{a}{Derived from MCMC analysis}
\tablenotetext{b}{Derived from spectral analysis}
\label{hostparams}
\end{deluxetable} 

\begin{deluxetable}{lr}
\tablewidth{0pt}
\tablecaption{Parameters of the planet and orbit}
\tablehead{\colhead{Parameter} & \colhead{Value}}
\startdata
Transit epoch (BJD), $T_{\rm C}$ & $2454584.69823\pm 0.00029$\\
Orbital period, $P$ (d) & $3.7520656\pm 0.0000028$\\
Transit duration, $T_{\rm 14}$ (d) & $0.1548\pm 0.0014$\\
Planet/star area ratio, $R_{\rm P}^{2}$/R$_{*}^{2}$ & $0.0099\pm 0.0002$\\
Impact parameter, $b$ (R$_{*}$) & 0.568$^{+ 0.038}_{- 0.046}$\\
Stellar reflex vel., K$_{\rm 1}$ (km s$^{-1}$) & $0.0634\pm 0.0038$\\
Centre-of-mass vel., $\gamma$ (km s$^{-1}$) & $-2.3439\pm0.0005$\\
Orbital separation, $a$ (AU) & $0.0499\pm 0.0018$\\
Orbital inclination, $i$ (deg) & $85.5\pm 0.5$\\
Orbital eccentricity, $e$ & $\equiv 0$ (adopted) \\

Planet mass, $M_{\rm p}$ (M$_{\rm J}$) & $0.542\pm 0.050$\\
Planet radius, $R_{\rm p}$ (R$_{\rm J}$) & $1.428\pm 0.077$\\
Planet surface gravity, $\log g_{\rm p}$ (cgs) & $2.784\pm 0.044$\\
Planet density, $\rho_{\rm p}$ ($\rho_{\rm J}$) & $0.186\pm 0.026$\\
Planet density, $\rho_{\rm p}$ (cgs) & $0.247\pm 0.035$\\
Planet equil. temp. (A=0), $T_{\rm p}$ (K) & $1652\pm 28$\\
\enddata
\label{sysparams}
\end{deluxetable} 

\section{System parameters and discussion}
The available WASP-South and EulerCAM photometric data were combined with the
CORALIE radial-velocity measurements in a simultaneous Markov-Chain
Monte-Carlo (MCMC) analysis, as described in \cite{cameron07}. An initial run
yielded a best-fit value for the orbital eccentricity nearly consistent with
zero ($e=0.052^{+0.029}_{-0.040}$), so a further fit was made with the
eccentricity fixed to zero. We chose not to excise from the fit any data
points in the EulerCAM light-curves affected by systematics; nevertheless the
fitted model light-curves (Fig.~\ref{foldlc}) globally do not appear to have
been adversely perturbed by these features.

The best-fit system parameters are listed in Tables~\ref {hostparams} and
~\ref{sysparams} and reveal WASP-15b to be a planet with one of the lowest
densities yet measured, comparable to TrES-4 \citep{mandushev07, sozzetti08}.
The planetary radius measured here lies above that predicted by the models of
\citet{fortney07} and \citet{burrows07} for a coreless planet of the mass, age
and insolation of WASP-15b. An additional internal heat source such as tidal
dissipation may be required in order to account for this anomalously large
radius. Indeed \citet{liu08} have recently shown that only moderate tidal
heating would be required to explain the radius anomalies of planets such as
TrES-4, and that the degree of heating is plausible if it is assumed that the
orbital eccentricities of such systems are non-zero yet still consistent with
the observational limits.


\begin{thebibliography}{}
\bibitem[Bakos et al.(2004)]{bakos04} Bakos, G., Noyes, R.~W., Kov{\'a}cs, G.,
  Stanek, K.~Z., Sasselov, D.~D., \& Domsa, I.\ 2004, \pasp, 116, 266

\bibitem[Barklem et al.(2000)]{barklem00} Barklem, P.~S., Piskunov,
  N., \& O'Mara, B.~J.\ 2000, \aaps, 142, 467 

\bibitem[B{\"o}hm-Vitense(2004)]{bohm04} B{\"o}hm-Vitense, E.\ 2004, \aj, 128, 2435 

\bibitem[Burrows et al.(2007)]{burrows07} Burrows, A., Hubeny, I., Budaj, J.,
  \& Hubbard, W.~B.\ 2007, \apj, 661, 502

\bibitem[Castelli et al.(1997)]{castelli97} Castelli, F., Gratton, R.~G., \&
  Kurucz, R.~L.\ 1997, \aap, 318, 841  

\bibitem[Collier Cameron et al.(2006)]{cameron06} Collier 
Cameron, A., et al.\ 2006, \mnras, 373, 799 

\bibitem[Collier Cameron et al.(2007)]{cameron07} Collier 
Cameron, A., et al.\ 2007, \mnras, 380, 1230 

\bibitem[Fortney et al.(2007)]{fortney07} Fortney, J.~J., Marley, 
M.~S., \& Barnes, J.~W.\ 2007, \apj, 659, 1661 

\bibitem[Girardi et al.(2000)]{girardi00} Girardi, L., Bressan, A., Bertelli,
  G., \& Chiosi, C.\ 2000, \aaps, 141, 371 

\bibitem[Gonzalez et al.(2001)]{gonzalez01} Gonzalez, G., Laws, 
C., Tyagi, S., \& Reddy, B.~E.\ 2001, \aj, 121, 432 

\bibitem[Gonzalez \& Laws(2000)]{gonzalez00} Gonzalez, G., \& Laws,
  C.\ 2000, \aj, 119, 390 

\bibitem[Kurucz \& Bell(1995)]{kurucz95} Kurucz, R., \& Bell, B.\
  1995, Atomic Line Data (R.L.~Kurucz and B.~Bell) Kurucz CD-ROM
  No.~23.~Cambridge, Mass.: Smithsonian Astrophysical Observatory, 1995., 23

\bibitem[Liu et al.(2008)]{liu08} Liu, X., Burrows, A., \& Ibgui, L.\ 2008,
  \apj, 687, 1191 

\bibitem[Magain(1984)]{magain84} Magain, P.\ 1984, \aap, 134, 189 

\bibitem[Mandushev et al.(2007)]{mandushev07} Mandushev, G., et 
al.\ 2007, \apjl, 667, L195 

\bibitem[McCullough et al.(2005)]{mccullough05} McCullough, P.~R., 
Stys, J.~E., Valenti, J.~A., Fleming, S.~W., Janes, K.~A., 
\& Heasley, J.~N.\ 2005, \pasp, 117, 783 

\bibitem[O'Donovan et al.(2006)]{odonovan06} O'Donovan, F.~T., Charbonneau,
  D., \& Hillenbrand, L.\ 2006, Bulletin of the American Astronomical Society,
  38, 1212

\bibitem[Pollacco et al.(2006)]{pollacco06} Pollacco, D.~L., et 
al.\ 2006, \pasp, 118, 1407 

\bibitem[Santos et al.(2004)]{santos04} Santos, N.~C., Israelian, G., \&
  Mayor, M.\ 2004, \aap, 415, 1153 

\bibitem[Smalley(2005)]{smalley05} Smalley, B.\ 2005, Memorie 
della Societa Astronomica Italiana Supplement, 8, 130 

\bibitem[Smalley(2001)]{smalley01} Smalley B., Smith K. C., Dworetsky M. M.,
  2001, UCLSYN Userguide 

\bibitem[Smith(1992)]{smith92} Smith K.C., 1992, PhD thesis, University of London

\bibitem[Sozzetti et al.(2008)]{sozzetti08} Sozzetti, A., et al.\ 2008, arXiv:0809.4589

\bibitem[Valenti 
\& Fischer(2005)]{valenti05} Valenti, J.~A., \& Fischer, D.~A.\ 2005, \apjs, 159, 141 

\end{thebibliography}
\end{document}